\documentclass[reprint,twocolumn,superscriptaddress,secnumarabic,amssymb, nobibnotes, aps, prl]{revtex4-1}

\usepackage{extarrows}
\usepackage{amsmath}    
\usepackage{graphicx}   
\usepackage{verbatim}   
\usepackage{color}      
\usepackage{subfigure}  
\usepackage{hyperref}   
\usepackage{verbatim}   
\newcommand{\bk}{\mathbf{k}}

\begin{document}
\title{Antiferromagnetism in RuO$_2$ as $d$-wave Pomeranchuk instability }

\author{Kyo-Hoon Ahn}
\affiliation{Division of Display and Semiconductor Physics, Korea University, Sejong 30019, Korea}
\author{Atsushi Hariki}
\affiliation{Institute for Solid State Physics, TU Wien, 1040 Vienna, Austria}
\author{Kwan-Woo Lee}
\affiliation{Division of Display and Semiconductor Physics, Korea University, Sejong 30019, Korea}
\affiliation{Department of Applied Physics, Graduate School, Korea University, Sejong 30019, Korea}
\author{Jan Kune\v{s}}
\affiliation{Institute for Solid State Physics, TU Wien, 1040 Vienna, Austria}
\affiliation{Institute of Physics, Czech Academy of Sciences, Na Slovance 2, 182 21 Praha 8, Czechia}
\date{\today}

\begin{abstract}
We present a computational study of antiferromagnetic transition in 
RuO$_2$. The rutile structure with the magnetic sublattices coupled by $\pi/2$-rotation leads to a spin-polarized band structure in the antiferromagnetic state, which gives rise to a $d$-wave modulation of the Fermi surface in the spin-triplet channel. We argue a finite spin conductivity  that changes sign in the $ab$ plane is expected RuO$_2$ because of this band structure. We analyze the
origin of the antiferromagnetic instability and link it to presence of a nodal line close to the Fermi level.
\end{abstract}

\maketitle
Antiferromagnetic (AFM) metals have been attracting much interest recently due to their spintronic applications based on coupling between magnetic moments and charge current~\cite{wadley2016}. 
While ubiquitous among $3d$ transition metal oxides, antiferromagnetism in this group is typical for Mott insulators rather than metals. Metallic antiferromagnets can be found either doping Mott insulators as in cuprates, by replacing oxygen with more covalent ligands as in iron pnictides~\cite{Kamihara2008} or {CuMnAs}~\cite{Wadley2013}, or by moving down the periodic table to less correlated $4d$ transition metals. 
Itinerant antiferromagnetism, with magnetism and transport governed by the same
electronic states, arises usually via the Slater mechanism~\cite{Slater1951}. Nesting between parts of the Fermi surface (FS) produces an instability that is resolved by an AFM state, which is
stabilized by gapping of the nested parts of FS. Such an AFM state 
reduces the translation symmetry of the system.
Its band structure is not spin polarized since the spin-up and -down sublattices are connected by a sublattice translation within the AFM unit cell. 

Structures with even number of magnetic atoms in the unit cell allow AFM ordering without breaking of translation symmetry. Ruthenium dioxide RuO$_2$ with rutile structure, recently observed to be a room temperature antiferromagnet~\cite{Berlijn2017,Zhu2019}, is an example of such a material. 
In this Letter we report a computational study of RuO$_2$ using density functional + dynamical mean-field theory (DFT+DMFT) as well as static Hartree-Fock (HF) techniques. We find that 
AFM order in RuO$_2$ leads to a spin-polarized band structure. The distinct spin-up and spin-down 
FSs are rotated by $\pi/2$ along the crystallographic $c$-axis with respect to each other. The AFM order thus can be viewed as a $d$-wave Pomeranchuk instability in the spin-triplet channel~\cite{Wu07}. We discuss the experimental implications and trace the origin of 
the AFM instability to a nodal line in the paramagnetic (PM) band structure located 
in the vicinity of Fermi level. 

\begin{figure}[t]
\centering{
\includegraphics[width=90mm]{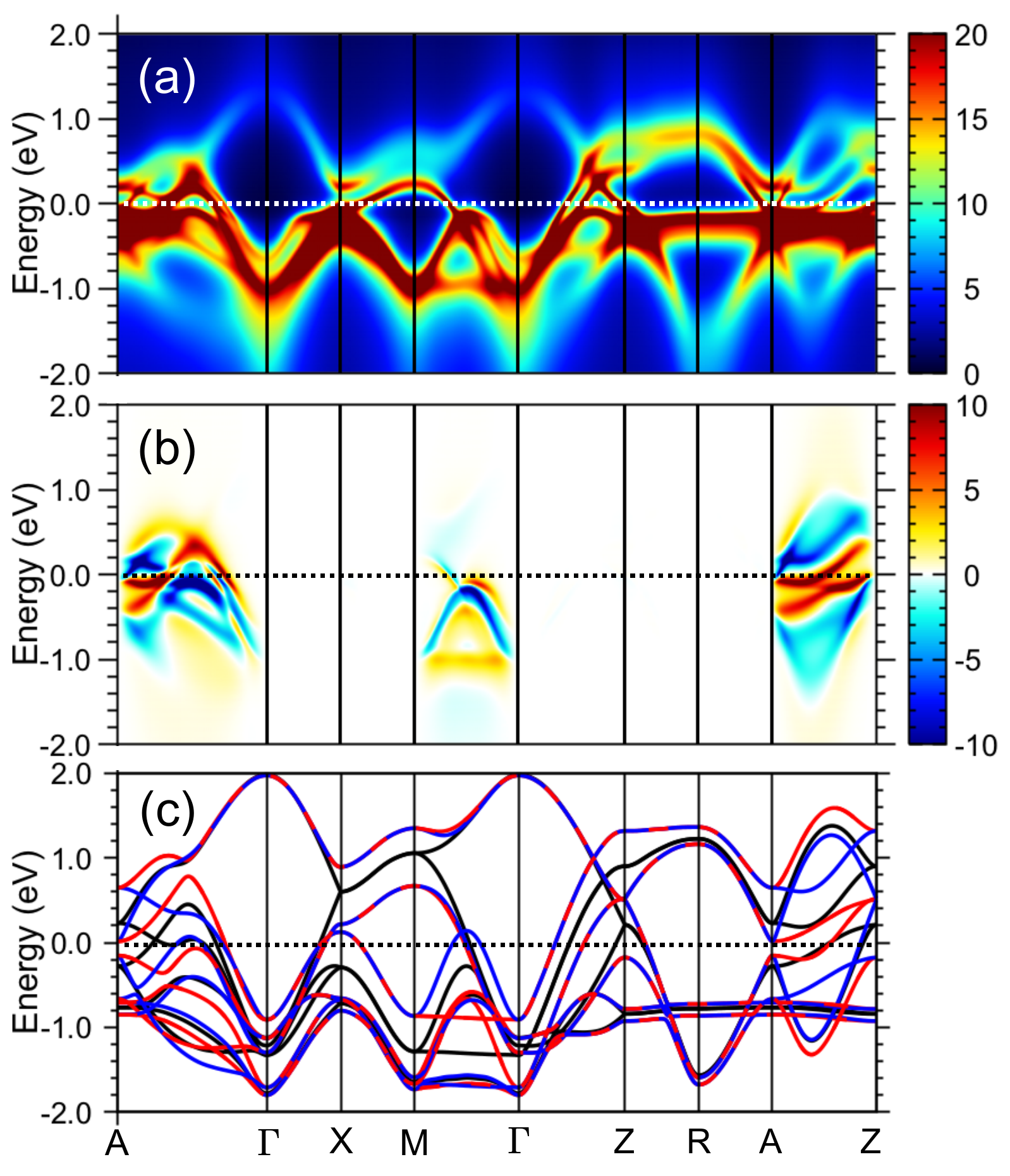}
}
\caption{(a) DFT+DMFT band structure, the spectral function $A_{\uparrow\uparrow}(\omega)+A_{\downarrow\downarrow}(\omega)$, along the high-symmetry lines
 in BZ. (b)
 The spin polarization of the band structure calculated as $A_{\uparrow\uparrow}(\omega)-A_{\downarrow\downarrow}(\omega)$
(c) HF band structure: paramagnetic (black) and AFM ($\uparrow$ red; $\downarrow$ blue).
The energy measured from the Fermi level.
}
\label{fig:bands}
\end{figure}

Starting from DFT electronic structure in the $P4_{2}/mnm$ structure~\cite{Berlijn2017} obtained with {\sc wien2k} package~\cite{wien2k}, 
we constructed Wannier orbitals spanning the Ru 4$d$ $t_{2g}$ bands
~\cite{wien2wannier,wannier90}.
The tight-binding model was augmented by intra-atomic electron-electron interaction of the Slater-Kanamori form~\cite{Slater1960,Kanamori1963}
\begin{equation*}
\begin{split}
H_U=&\sum_{\alpha}n_{\alpha\uparrow}n_{\alpha\downarrow}+\sum_{\alpha>\beta,\sigma\sigma'}  
(U-2J-\delta_{\sigma\sigma'})n_{\alpha\sigma}n_{\beta\sigma'}\\
+&\gamma J\sum_{\alpha\neq\beta}
\left(
c^{\dagger}_{\alpha\uparrow}c^{\dagger}_{\beta\downarrow}
c^{\phantom\dagger}_{\alpha\downarrow}c^{\phantom\dagger}_{\beta\uparrow}+
c^{\dagger}_{\alpha\uparrow}c^{\dagger}_{\alpha\downarrow}
c^{\phantom\dagger}_{\beta\downarrow}c^{\phantom\dagger}_{\beta\uparrow}+H.c.\right),
\\
\end{split}
\end{equation*}
where the indices $\alpha$, $\beta$ run over the orbital flavors on the same atom, whose
site index was dropped for simplicity. The resulting Hubbard model was studied with DMFT and HF methods. 
The DMFT calculations employed the strong-coupling continuous-time quantum Monte Carlo method~\cite{Werner2006a,Boehnke2011,Hafermann12,Hariki2015} and the density-density approximation ($\gamma=0 $) to $H_U$. HF calculations with spin-orbit coupling (SOC), included {\it a posteriori}
to the mean-field Hamiltonian, were performed to determine the orientation of the local moments
in the AFM state. The impact of SOC on the band structure was found to be only minor in line
with the conclusions of Ref.~\onlinecite{Jovic2018}.

The calculations covered a wide range of $U$, $J$ 
parameters and temperatures, summarized in the Supplemental Material (SM)~\cite{SM}. While size of the ordered moment varies with these parameters and among the two methods the general observations presented here are shared by all calculations. Quantitatively, the DMFT and HF electronic structures 
corresponding to the same ordered moment agree with each other and as well as with DFT+U calculations, as shown in SM~\cite{SM}. This observation suggests that staggered Weiss field due to the AFM order, a feature shared by all methods, is the dominant effect.
In the following, we present HF (DMFT) results for $U=1.7$ (1.7)~eV and $J=0.2$ (0.45)~eV at $T=300$ (100)~K.


{\it Spin-polarized band structure.} In Fig.~\ref{fig:bands}a we show the AFM band structure obtained with DMFT. The band structure throughout the Brillouin zone (BZ) is spin-polarized, 
giving rise to a spin contrast shown in Fig.~\ref{fig:bands}b. Special
high-symmetry planes, discussed below, are an exception where spin-up and spin-down bands are degenerate.
In Fig.~\ref{fig:bands}c we present PM and AFM bands obtained with HF method. The ordered moment in the Wannier basis is $\sim0.8$~$\mu_B$ in both cases. Apart from noticeable dynamical bandwidth renormalization absent in the HF spectra, we observe overall a good agreement between the HF and DMFT results. This suggests that the weak-coupling HF approach provides a reasonable description of the physics of RuO$_2$ and we can use its simple structure to analyze the observed behavior.

Since both Ru$_1$ and Ru$_2$ sites with the anti-parallel spin moments fit in the rutile unit cell,
the AFM order does not affect the translation symmetry. It reduces the point symmetry, however. The operations connecting the magnetic sublattices (mapping Ru$_1$ to Ru$_2$), e.g., the $4_2$ screw rotation, belong no more to the symmetry group. To do so they must be augmented with spin-inversion. 
As a result, the AFM band structure is spin-polarized with the two spin channels being coupled by $\pi/2$-rotation. This distinguishes RuO$_2$ from other antiferromagnets with spin-polarized bands, such as tetragonal CuMnAs~\cite{Wadley2013} where the magnetic sublattices, and thus the spin-polarized bands, are connected by inversion symmetry. The spin-up and spin-down bands of RuO$_2$ are degenerate (cross) along the $k_a=0,\pi$ planes, which are invariant under the glide plane $x_a=0$ connecting 
the magnetic sublattices. The same applies for $k_b=0,\pi$ planes and $x_b=0$ glide plane.

\begin{figure}[t]
\centering{
\includegraphics[scale=0.65]{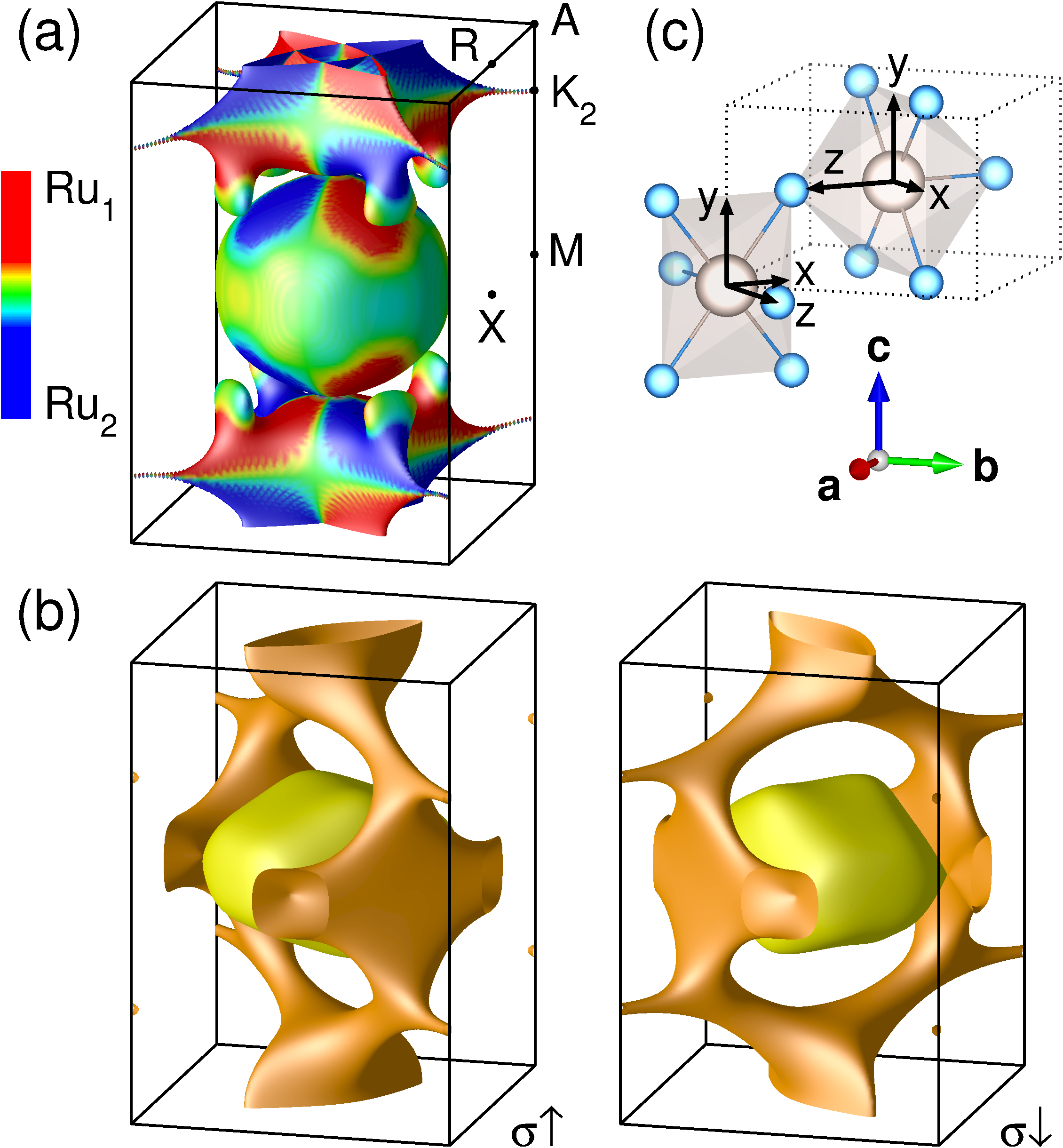}
}
\caption{(a) Fermi surface of paramagnetic RuO$_2$ obtained with HF method. The color codes the 
atomic polarization (difference of the Ru$_1$ and Ru$_2$ contributions to the wavefunction at a given k-point).
(b) Spin-polarized Fermi surface of the AFM state.
(c) Crystal structure of RuO$_2$ with global, $\mathbf{a}$, $\mathbf{b}$, $\mathbf{c}$,
and local coordinates indicated.
}
\label{fig:fermi}
\end{figure}

\begin{figure}[t]
\centering{
\includegraphics[width=83mm]{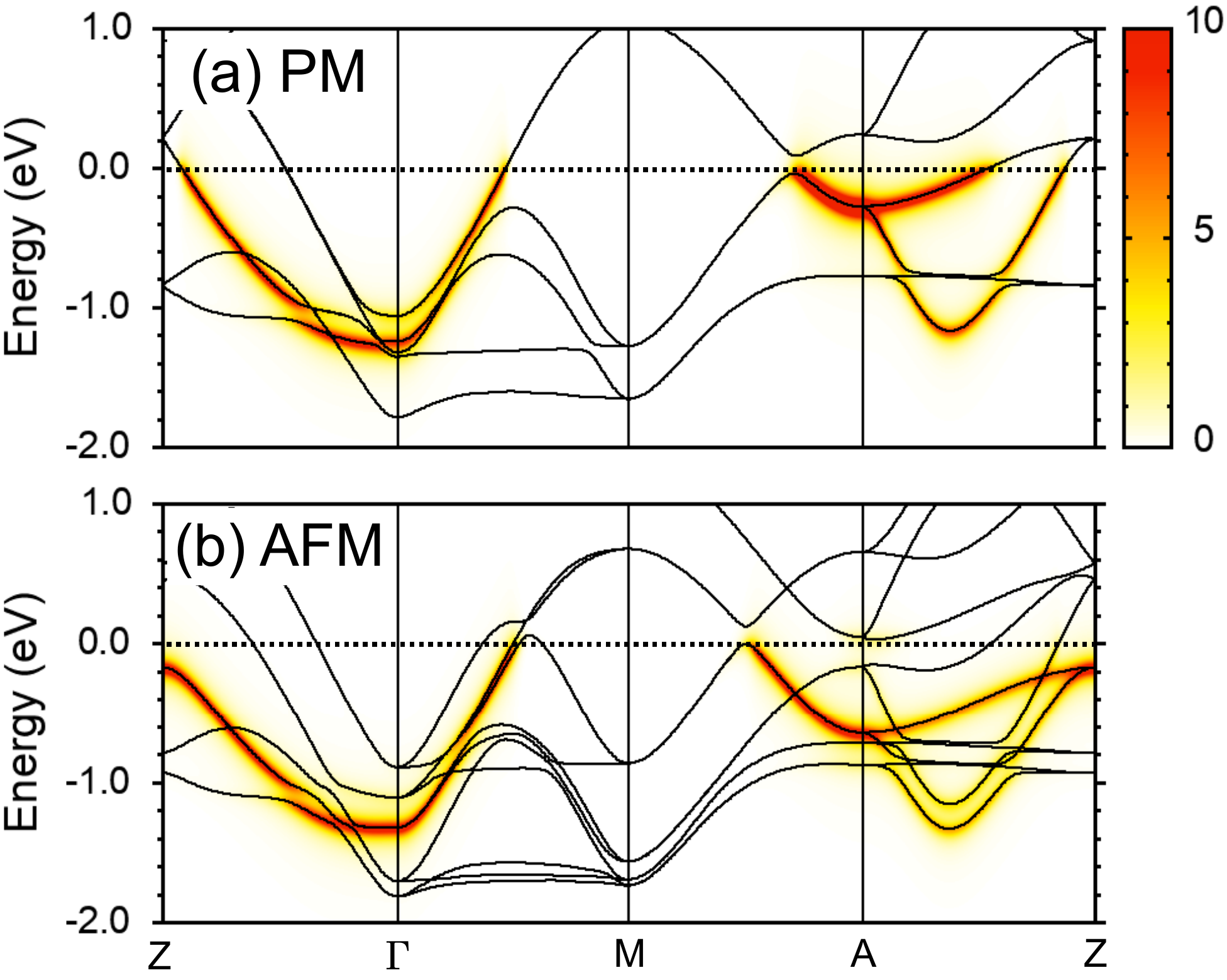}
}
\caption{(a) Paramagnetic and (b) antiferromagnetic band structure obtained with the HF method. Here the SOC is included. The color map shows simulated ARPES spectra. Lorentzian broadening of half-width 0.05~eV and Fermi-Dirac distribution at 300~K are included.}
\label{fig:arpes}
\end{figure}

The spin polarized band structure gives rise to FS shown
in Fig.~\ref{fig:fermi}b. The deformation of four-fold symmetric FS
in Fig.~\ref{fig:fermi}a into a pair of two-fold symmetric FS connected
by $\pi/2$-rotation and spin inversion can be classified as
$d$-wave spin-triplet Pomeranchuk instability of the $\alpha$ 
type in the notation of Ref.~\cite{Wu07}. This type of spin-polarized FS
implies a finite longitudinal spin conductivity $\sigma_{\uparrow}-\sigma_{\downarrow}$ in the $ab$-plane with opposite sign of in the (1,1,0) and (1,-1,0) directions, which vanishes
along the crystallographic axes.

Apart from spin-polarization the AFM order causes changes of the band structure that can be detected with angle-resolved photoemission spectroscopy (ARPES). In Ref.~\cite{Jovic2018}, soft x-ray ARPES spectra obtained with $p$-polarized light were reported. We have simulated $p$-polarized ARPES spectra, assuming that the scattering plane is perpendicular to the $c$-axis. The resulting spectra in the PM and AFM states are shown in Fig.~\ref{fig:arpes}. The spectra along $AM$ and $M\Gamma$ directions agree well with those of Fig.~S3 of Ref.~\onlinecite{Jovic2018}, while the weaker signal observed observed halfway between $\Gamma$ and $Z$ should be silent according to our simulation~\footnote{This discrepancy could arise from complexities of the matrix-element effects (e.g.~photon energy dependence) in ARPES spectra or surface effects. In the present simulation, the selection rule for the orbital excitation under the scattering plane perpendicular to the $c$-axis and $p$-polarized x-rays is taken into account.}. Comparison of the PM and AFM bands reveals a sizable shift of the FS crossing point on $MA$ with the onset of AFM order, while the crossing point on $\Gamma M$ moves only slightly. The sharp experimental band observed along $MA$ line does not produce any spin contrast. On the other hand well separated spin polarized bands predicted along the $AZ$ line by the present calculation have vanishing cross section for
the $p$ polarization of incoming photons.

\begin{figure}[t]
\centering{
\includegraphics[scale=0.14]{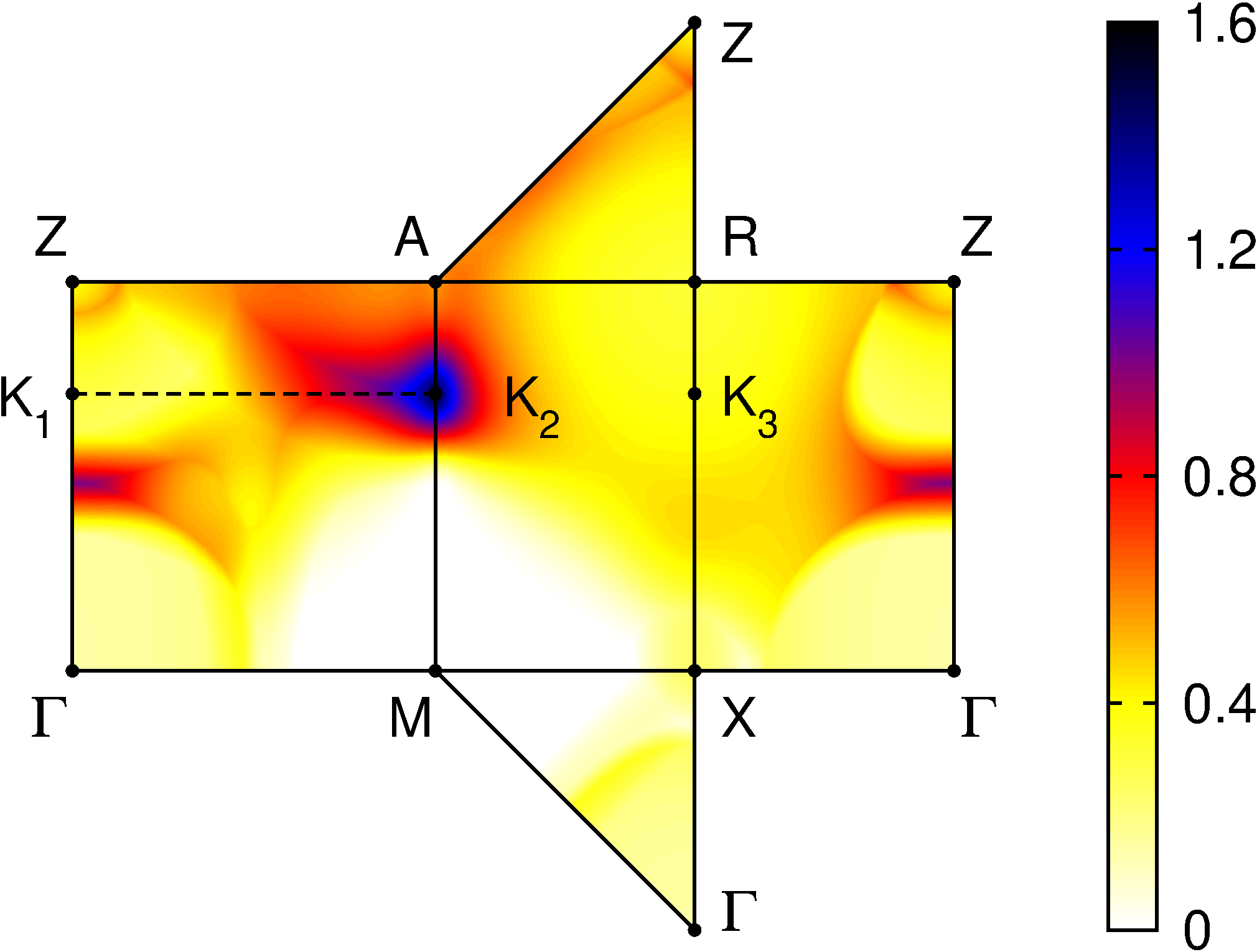}
}
\caption{The contribution $\eta(k)$ (defined in the text) to the condensation energy  along several cuts through BZ. 
}
\label{fig:diff}
\end{figure}

\begin{figure}[t]
\centering
\includegraphics[scale=0.3]{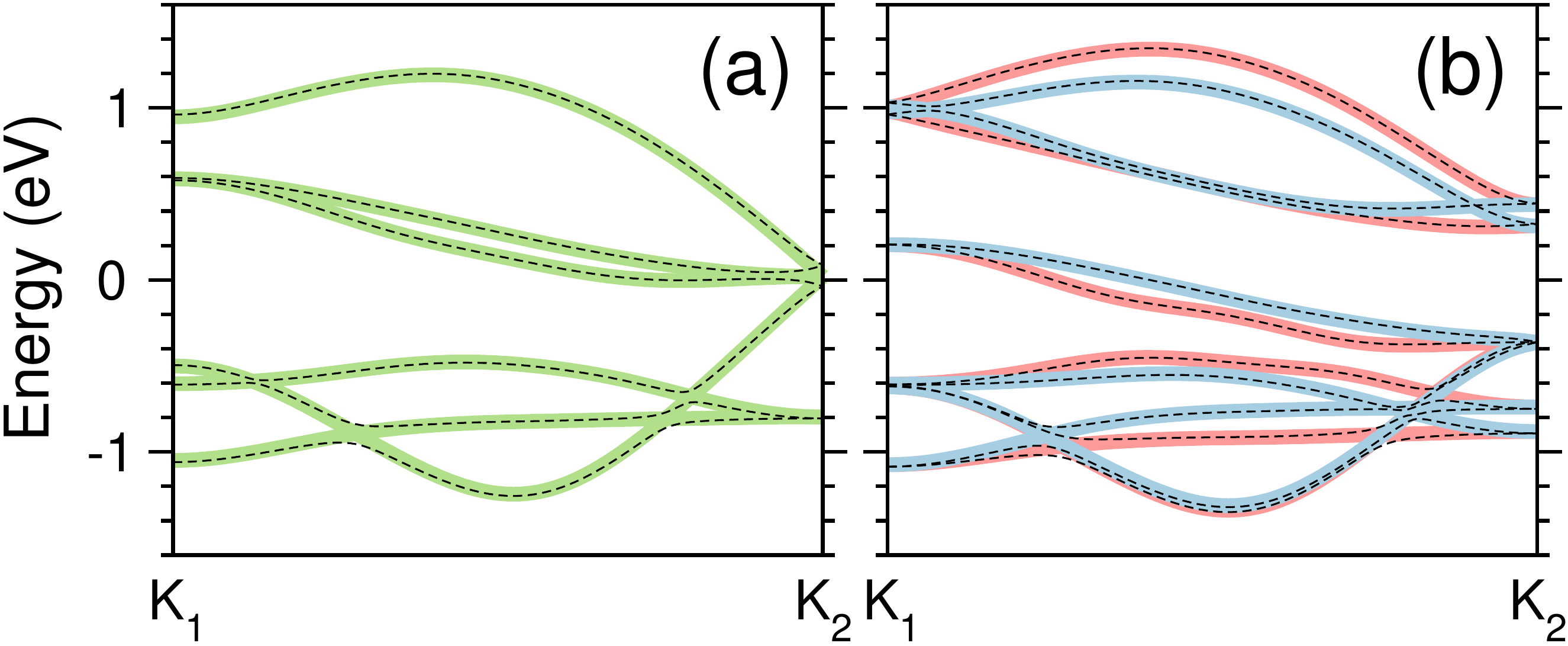}
\includegraphics[scale=0.11]{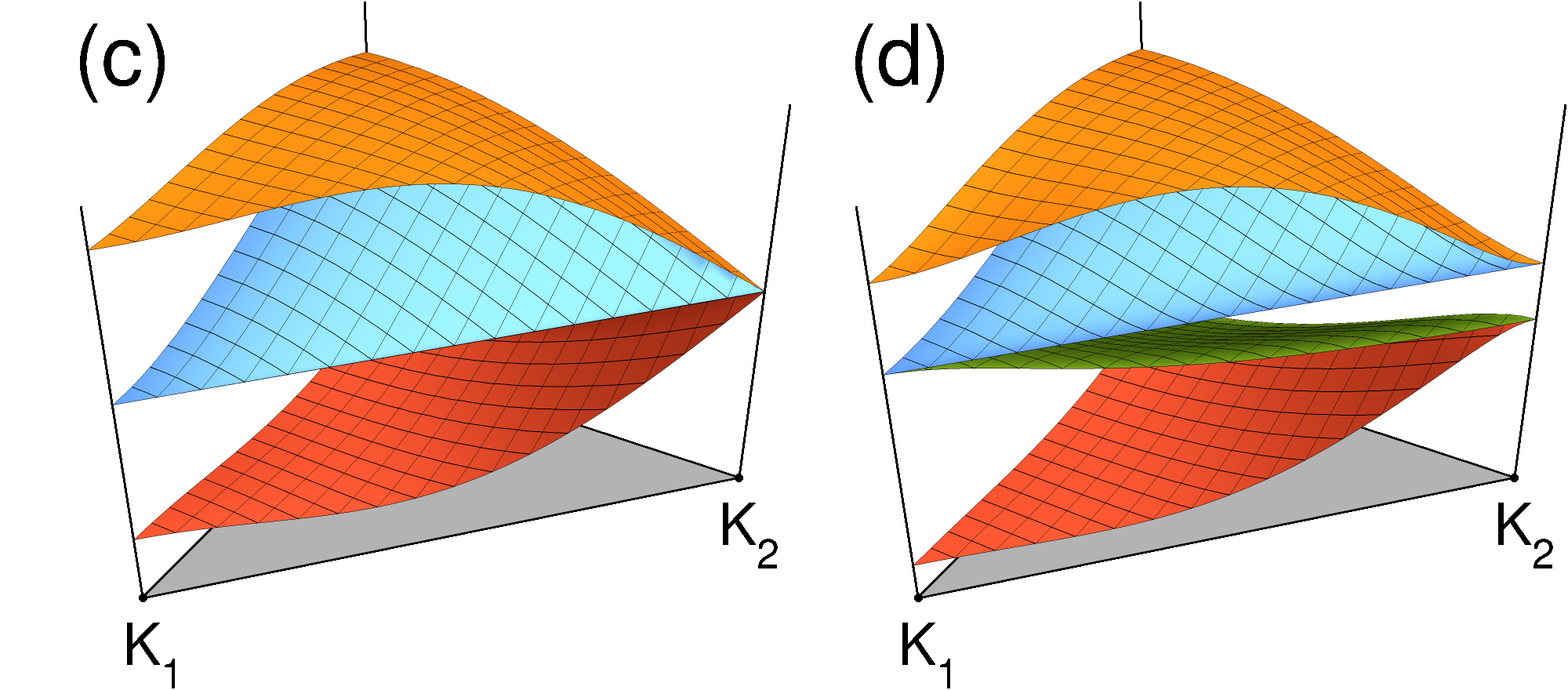}
\caption{Bandstructures along $K_1K_2$ line of the HF model with (a) the paramagnetic (green) and (b) the AFM ($\uparrow$ red; $\downarrow$ blue) phase. The effect of SOC is represented with dashed (black) lines. (c) Paramagnetic bandstructure in the
horizontal $K_1K_2K_3$ plane obtained with dominant nn hopping. (d) The same as in (c) with a staggered potential.
}
\label{fig:band_detail}
\end{figure}

{\it Origin of AFM instability.} 
Similarity between the DMFT and HF results allows us to use
the simpler HF approach to analyze the origin of AFM instability.
The grand potential in the HF approach with mean-field $\Delta$ is given by the sum 
of the eigenenergies $\epsilon_{{nk,\sigma}}^\Delta$ of occupied single-particle states of the mean-field Hamiltonian (measured from the chemical potential) and 
a positive $\Delta$-dependent constant
$\langle \Omega_{\text{MF}}\rangle=\sum_{nk,\sigma}\epsilon_{{nk,\sigma}}^\Delta f_{nk,\sigma}+C(\Delta)$,
where $f_{nk,\sigma}$ is the occupation number.
The AFM state is stabilized if lowering 
of the first term overcomes the increase of the second one relative to the PM solution.
To assess how different parts of BZ contribute to stabilization 
of the AFM order we plot the difference ${\eta(k)=\sum_{n,\sigma}(\epsilon^{PM}_{nk,\sigma}-\epsilon^{AFM}_{nk,\sigma})
f_{nk,\sigma}}$ in Fig.~\ref{fig:diff}. We find a hot spot around the point marked $K_2$, which extends horizontally towards
the zone center. This observation contrasts with the result of Ref.~\cite{Berlijn2017} where a point on the $RX$ line was identified as a hot spot destabilizing
the PM phase. Large contributions to condensation energy are expected from regions where gaps open at the Fermi level. This is the case of the vicinity of $K_2$, analyzed in Fig.~\ref{fig:band_detail}. 

At the point $K_2$ the nodal line NDL1 of Ref.~\cite{Jovic2018} reaches the edge of BZ. Together with the band sticking along vertical faces of BZ, this gives rise to a four-fold degenerate point, which happens to be very close to the Fermi level.
The band structure in the vicinity of $MA$ line has a simple explanation in terms of
inter-atomic hopping. The relevant electronic states are formed by $zx$ and $yz$ orbitals
in the local coordinates shown in Fig.~\ref{fig:fermi}a. Along the $MA$ line the two Ru sites as well as $zx$ and $yz$ orbitals are decoupled and the dispersion is governed by
hopping along chains of edge sharing RuO$_6$ octahedra stretching along the $c$-axis. The opposite sign of $zx$-$zx$ and $yz$-$yz$ hopping results in the crossing of the corresponding bands at $K_2$. The different parity of the $zx$ and $yz$ bands with respect to $c$-axis inversion 
explains their appearance/absence in the ARPES spectrum of Ref.~\cite{Jovic2018}.

The dispersion in $K_1K_2K_3$ plane is governed by the inter-sublattice Ru$_1$-Ru$_2$ hopping
(see SM~\cite{SM} for details). The nodal line NDL1 that is flat connecting $K_2$ and $K_2$ points in the $nn$-approximations, Fig.~\ref{fig:band_detail}c, acquires some corrugation when long-range hopping and $x^2-y^2$ orbitals are included, see Fig.~\ref{fig:band_detail}a and
Ref.~\cite{Jovic2018}. The AFM order introduces a staggered potential with opposite sign in each spin channel. Its effect on HF bands along $K_1K_2$ line is shown in Fig.~\ref{fig:band_detail}b (both spin channels) and on the model bands in Fig.~\ref{fig:band_detail}d (one spin channel).
The gap opening by the staggered potential can be illustrated by expanding 
the tight-binding Hamiltonian to linear order around $K_2=(\pi,\pi,z)$
\begin{equation}
\label{eq:hlin}
\begin{split}
&h(\bk)=\\
&\begin{pmatrix} 
\Delta +\alpha k_c & 0 & 0 & \gamma(k_a-k_b) \\
0 & \Delta -\beta k_c & \gamma (k_a+k_b) & 0 \\
0 & \gamma^*(k_a+k_b) & -\Delta + \alpha k_c& 0 \\
\gamma^* (k_a-k_b) & 0 & 0 & -\Delta-\beta k_c
\end{pmatrix},
\end{split}
\end{equation}
in the basis formed by $yx(1)$, $xz(1)$, $yz(2)$, and $xz(2)$ orbitals. 

To open a gap by the staggered potential $\Delta$
we need a (approximate) band crossing close to the Fermi level and hybridization between the magnetic sublattices (Ru$_1$-Ru$_2$) so that the band crossing is not just shifted to a different position in BZ. Such a band structure, described by Eq.~\ref{eq:hlin}, is found in the vicinity of NDL1.
In the PM state such $k$-regions are characterized by rapid changes in the Ru$_1$/Ru$_2$ sublattice composition of the wave functions.  In Fig.~\ref{fig:fermi}a we show the PM FS colored by the sublattice polarization
(difference between Ru$_1$ and Ru$_2$ weight) of the corresponding wave functions. The hot spots in Fig.~\ref{fig:diff} correlate with red/blue boundaries in Fig.~\ref{fig:fermi}a where sublattice polarized bands
meet. The sublattice polarization descends to a spin polarization in the AFM phase, Fig.~\ref{fig:fermi}b. On the other hand, the regions with fifty-fifty
sublattice participation are insensitive to the staggered potential, e.g. in the $\Gamma XM$ plane, and PM and AFM FS essentially coincide with each other.

Finally, we discuss the role of SOC. The band structure calculations of Ref.~\cite{Jovic2018} found a minor modifications of the band structure in the form of avoided band crossings. We find that gapping NDL1 due to SOC has only minor effect on the AFM instability. The impact on the HF band structure depends strongly on the type of mean-field decoupling, a deficiency that is not present in DMFT treatment. 
In the present work we have added SOC
{\it a posteriori} to the converged HF, which has the similar effect as in DFT calculation~\cite{Jovic2018}. The band structures  with and without SOC are compared in Fig.~\ref{fig:band_detail}. The main effect of SOC is to break the spin isotropy reflected in magneto-crystalline anisotropy. The HF+SOC calculations yield Ru moments parallel to the $c$-axis.

The rutile structure of RuO$_2$ gives rise to several interesting phenomena. \v{S}mejkal {\it et al.}~\cite{Smejkal2019} recently suggested realization of crystalline Hall effect in this material. The $d$-wave modulation of FS results in a finite longitudinal spin conductivity with a sign change between (1,1,0) and (1,-1,0) directions. Another interesting question is response of the AFM structure to external magnetic field. Using the weak-coupling approach the authors of Ref.~\onlinecite{Wu07}
concluded that the order parameter (staggered moment in this case) aligns parallel to the external field. This is associated with an expansion of FS for the parallel spin component and shrinking of the anti-parallel one, with the consequence of breaking the four-fold symmetry in the charge channel.
Such behavior of an antiferromagnet would be rather unusual and require extremely
soft magnetic moments. We have investigated this possibility with HF calculations (without SOC), but found the conventional behavior with moments turning perpendicular to the external field with a small tilt into the field direction, in which case the $\pi/2$ symmetry between the spin-up and spin-down FS
is reserved.

In conclusion, we have studied the antiferromagnetism of RuO$_2$ using combinations
of DFT band structure with HF and DMFT treatment of intra-atomic interaction.
The AFM ordering in RuO$_2$ can be classified as a spin-triplet $d$-wave Pomeranchuk instability of FS. 
It gives rise to a spin-polarized band structure that leads to anisotropic spin-conductivity in the 
$ab$-plane. The source of AFM instability was traced to a nodal line that accidentally appears close to the Fermi level.

The authors thank W. E. Pickett,  J.~\v{Z}elezn\'y and K.~Yamagami for valuable discussions.
K.H.A. and K.W.L are supported by National Research Foundation (NRF) of Korea Grant No. NRF-2016R1A2B4009579.
A.H. and J.K. are supported by the European Research Council (ERC) under the European Union’s Horizon 2020 research and innovation programme (Grant Agreement No. 646807-EXMAG).
Access to computing and storage facilities provided by the Vienna Scientific Cluster (VSC) is greatly appreciated.

\end{document}